\documentclass[pra,twocolumn]{revtex4}
\usepackage{graphicx}
\begin{document}

\title{Generating qudits with $d=3,4$ encoded on two-photon states}
\author{Giacomo Mauro D'Ariano$^{a,c}$, Paolo Mataloni$^{b,c}$, and
Massimiliano F. Sacchi$^{a,c}$} 

\affiliation{$^a$QUIT, Unit\`a INFM and
Dipartimento di Fisica ``A. Volta'', Universit\`a di Pavia, I-27100
Pavia, Italy}  

\affiliation{$^b$Dipartimento di Fisica,
Universit\`{a} di Roma `La Sapienza', I-00185 Roma, Italy}

\affiliation{$^c$CNISM, Consorzio Nazionale Interuniversitario per le Scienze
  Fisiche della Materia, Italy}

\begin{abstract}
We present an experimental method to engineer arbitrary pure states of
qudits with $d=3,4$ using linear optics and a single nonlinear
crystal.
\end{abstract}

\maketitle

Many issues in quantum information theory and processing deal with
qudits, namely $d$-level quantum systems, instead of qubits. The
interest in such more complex systems is both theoretical---the
general structure of quantum protocols can be simplified for arbitrary
dimension---and practical---some relevant applications perform better
using qudits. For example, new quantum cryptographic protocols were
recently proposed that deal specifically with qutrits
\cite{peres:00,kasz:03,lang:04} and the eavesdropping analysis showed
that these systems are more robust against specific classes of
eavesdropping attacks \cite{bruss:02,eav,durt:03,lang:04}. A further
advantage of using multilevel systems deals with novel fundamental
tests of quantum mechanics \cite{collins:02,zuk}.

Recent experimental realizations of qutrits rely on different physical
implementations. In interferometric schemes, qutrits are generated by
sending an entangled photon pair through a multi-armed interferometer
\cite{rob:01}, and the number of arms defines the dimensionality of
the system. Other techniques exploit the properties of orbital angular
momentum of single photons \cite{lang:04,zei:01,zei:03}, or perform
postselection from four-photon states \cite{antia:01}. All the above
techniques, however, provide only partial control over the qutrit
state. In the method of Refs. \cite{lang:04,zei:01,zei:03} one needs a
specific hologram for a given qutrit state. Also the interferometric
scheme \cite{rob:01} is not very flexible in switching between
different states. More recently, an experimental realization of
arbitrary qutrit states has been reported \cite{rece}, where the
polarization state of two-photon field has been exploited. Such a
realization requires the use of \emph{three} nonlinear crystals pumped
by a common coherent source.

In this paper we show an experimental method to engineer arbitrary
pure states of qutrits and ququads, using a single nonlinear crystal
and linear optical devices as phase waveplates. The qudit is encoded
on the polarization of a two-photon state, and is obtained from local
(e.g.  single-photon) unitary transformations on a pure non-maximally
entangled state which plays the role of a\emph{\ seed} state. It can
be generated a parametric source of entangled photon states
\cite{note,roma1}. In the present paper we refer to a high brilliance
source \cite{roma1}, with high flexibility in terms of state
generation. It has been recently demonstrated that by this it is
possible to produce two photon hyper-entangled states, entangled in
polarization and momentum \cite{roma2}.  Indeed, the adoption of
hyper-entangled states can be crucial whenever one is interested in
quantum information applications of qudits, since hyper-entanglement
in polarization and momentum allows one to perform non trivial
measurements---such as Bell measurements \cite{padua}---which are
needed for quantum key distribution. In fact, as we will show, it is
possible to implement a quantum cryptographic scheme with ququads that
exploits two mutually unbiased bases made by two-photon Bell states,
and here hyper-entanglement allows to perform Bell measurements.

In the following we first show how to obtain an arbitrary qudit with
$d=3,4$, from local unitary transformations on a bipartite pure state
of two qubits. Hence, we want to show how to generate a state of the
form
\[
|\psi \rangle = \alpha |00\rangle +\beta |11\rangle +\gamma |01\rangle +\delta
|10\rangle \; 
\]
from the seed state 
\[
|\chi \rangle =x|00\rangle +\sqrt{1-x^{2}}|11\rangle \text{,} 
\]
by means of two local unitary transformations. In the state $|\psi
\rangle $ we can fix $\alpha $ positive, and take $\beta ,\gamma $ and
$\delta $ complex without loss of generality. The state $|\chi \rangle
$ is chosen with $x$ positive. Hence, given $\alpha ,\beta ,\gamma $
and $\delta $ we want to find $x$ and two unitaries $U$ and $W$ such
that
\begin{eqnarray}
|\psi \rangle =U\otimes W|\chi \rangle \;. 
\;\label{tre}
\end{eqnarray}
Of course, $x$, $U$, and $W$ will depend on the desired parameters
$\alpha ,\beta ,\gamma $ and $\delta $.

We can solve this problem by means of the singular value decomposition
(SVD), which states that for any matrix $A$ one can find two unitaries
$U$ and $W$ such that \cite{bhatia}
\begin{eqnarray}
A=U D W^{\tau }\;,\label{udw}
\end{eqnarray}
where $\tau $ denotes transposition on the fixed basis, and $D$ is
diagonal and positive.

Consider now the matrix $\Psi $ corresponding to the state $|\psi
\rangle $
\begin{eqnarray}
\Psi = \alpha  |0 \rangle 
\langle 0 | + \beta |1 \rangle \langle 1 | + \gamma |0
\rangle \langle 1 |+ \delta |1 \rangle \langle 0 | \;,
\end{eqnarray}
through the identity \cite{pla}
\begin{eqnarray}
|\psi \rangle = (\Psi \otimes I) (|00 \rangle + |11 \rangle )\;.
\end{eqnarray}
From the SVD $\Psi = U D W^\tau $ it follows that 
\begin{eqnarray}
|\psi \rangle &= &(U D W^\tau \otimes I)(|00 \rangle + |11 \rangle ) 
\nonumber \\
&= & (U D \otimes W)(|00 \rangle + |11 \rangle )  \nonumber \\
&= & (U \otimes W) (D\otimes I)(|00 \rangle + |11 \rangle )  \nonumber \\
&= & (U \otimes W) (d_1 |00 \rangle + d_2 |11 \rangle ) \;,
\end{eqnarray}
which is equivalent to Eq. (\ref{tre}). The values $d_1 $ and $d_2$
are the elements of the diagonal matrix $D$ (``the singular values of
$\Psi $'').  Notice that
\begin{eqnarray}
1=\langle \psi |\psi \rangle = \hbox{Tr}[\Psi ^\dag \Psi ]=\hbox{Tr}
[D^2]=d_1^2 +d_2^2 \;,
\end{eqnarray}
namely one obtains the correct normalization for the state $|\chi
\rangle $. Our result generally holds in arbitrary Hilbert spaces, and
hence provides a way to encode a qudit on a bipartite quantum system
of ${\cal H}\otimes {\cal H}$, by means of local unitary
transformations, where $d= (\hbox{dim}({\cal H}))^2$. Notice also that
the decomposition in Eq. (\ref{udw}) is not unique, and hence the
unitaries $U$ and $W$ in Eq. (\ref{tre}) are not unique either. For
example, one has the invariance property $U'=U Z$ and $W'=W Z^\dag $,
where $Z$ is an arbitrary diagonal unitary matrix.

Let us apply the above derivation to the case where the qubits are
represented by the polarization state of two photons. The seed state
is written
\begin{eqnarray}
|\chi \rangle =x|HH\rangle +\sqrt{1-x^{2}}|VV\rangle \;.
\label{chid}
\end{eqnarray}
The state in Eq. (\ref{chid}) represents a non-maximally entangled
polarization state. It is easily obtained from the source sketched in
Fig.~1.  It is based on a high stability single arm interferometer
which accomplishes the generation of the polarization entangled state
$|\Phi \rangle = \frac{1}{\sqrt{2}}\left( |H\rangle |H\rangle
+e^{i\theta }|V\rangle |V\rangle \right) $ by the superposition of the
degenerate parametric emission cones at wavelength $\lambda $ (see
Fig.~ 1) of a type-I \ $\beta $-$BaB_{2}O_{4}$ (BBO) crystal, excited
in two opposite directions, by a $V$-polarized laser beam at
wavelength $\lambda _{p}=\lambda /2$. Other basic elements of the
source are the following:

[$i$] A spherical mirror $M$, reflecting both the parametric radiation
or the pump beam, whose micrometric displacement allows to control the
state phase $\theta $ ($0\leq \theta \leq \pi $).

[$ii$] A zero-order $\lambda /4$ waveplate (wp), placed within the
$M-$BBO path, which performs the $|HH\rangle \rightarrow |VV\rangle $
transformation on the two-photon state belonging to the left-cone.

[$iii$] A positive lens which transforms the conical parametric
emission of the crystal into a cylindrical one, whose transverse
circular section identifies the so-called ''entanglement-ring''
($e-r$).

A zero-order $\lambda _{p}/4$ wp inserted between $M$ and the BBO
crystal, intercepting only the laser beam allows the engineering of
tunable non-maximally entangled states in the following way: the
polarization of the back-reflected pump beam is rotated by an angle
$2\theta _{p}$ with respect to the optical axis of the crystal when
the pump wp is rotated by an angle $ \theta _{p}$. As a consequence
the emission efficiency of the $\left | HH \right \rangle $
contribution is lowered by a coefficient proportional to $\cos
^{2}2\theta _{p}$, with $\theta _{p}$ adjusted in the range $0-\pi
/4$.  Alternatively, we can obtain a lower value of the $\left|
VV\right\rangle $ contribution with respect $\left| HH\right\rangle $
by inserting a $\lambda _{p}/2$ wp on the laser beam path before the
crystal. By simultaneous rotation of the two wp's, the complete
tunability of the entanglement degree can be achieved \cite{roma3}.

\begin{figure}[htb]
\begin{center}
\includegraphics[scale=.32]{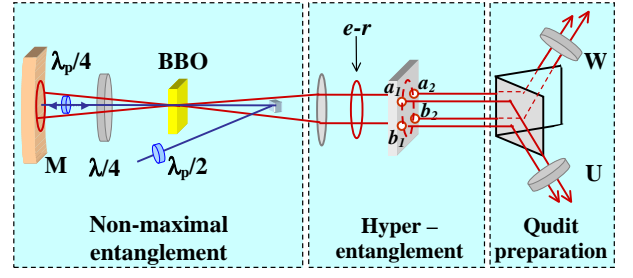}
\caption{Layout of the universal source of non-maximally polarization
entangled and polarization-momentum hyper-entangled two-photon
states. In the left part, non-maximally entangled states
in polarization are generated. In the central part, after division of
the entanglement ring ($e-r$) along a vertical axis by a prism-like
two-mirror system, momentum entanglement is realized by a four hole
screen which selects the correlated pairs of modes
$a_{1},b_{2}$ and $a_{2},b_{1}$. In the right part, qudits are encoded
by means of the local unitary transformations $ U$ and $W$ on modes
$a_{1},b_{1}$ and $a_{2},b_{2}$, respectively.}
\label{f:fig1}
\end{center}
\end{figure}

The local unitary transformations that are needed to generate the
desired state of the qudit can be easily realized by linear optics. In
fact, a unitary $2 \times 2$ matrix can generally be written as
\begin{eqnarray}
U=\left( 
\begin{array}{rr}
e^{i \alpha } \cos\theta
& e^{i \beta  }\sin\theta
\\ 
-e^{i \gamma } \sin\theta 
& e^{i(\beta + \gamma - \alpha )}\cos\theta
\end{array}
\right) 
\; 
\end{eqnarray}
Such a unitary can be factorized as follows
\begin{eqnarray}
U&=&\left( 
\begin{array}{lc}
e^{i \beta } & 0\\ 
0 & e^{i(\beta + \gamma - \alpha )}
\end{array}
\right) 
\left( 
\begin{array}{rl}
\cos\theta & \sin\theta\\ 
-\sin\theta  & \cos\theta
\end{array}
\right) 
\nonumber \\&  \times &
\left( 
\begin{array}{cl}
e^{i (\alpha -\beta )} & 0\\ 
0 & 1
\end{array}
\right) 
\;.\label{fact}
\end{eqnarray}
Hence, any unitary transformation on the polarization state of a
photon can be obtained as a sequence of a phase-shift, a rotation of
the polarization, and a final phase-shift.

\par The general scheme can be used to engineer mutually unbiased
bases \cite{mub} of qutrits for cryptographic purposes. For example,
the basis
\begin{eqnarray}
&&|u_I \rangle = |HH \rangle \;,  \nonumber 
\\
& & |u_{II} \rangle = |VV \rangle \;,  \nonumber \\
& & |u_{III} \rangle = \frac {1}{\sqrt 2}(|HV \rangle + |VH \rangle )\equiv
|\psi ^+ \rangle \;;\label{u3}
\end{eqnarray}
is mutually unbiased with the basis 
\begin{eqnarray}
&&|v_{I} \rangle = \frac {1}{\sqrt 3} \left (|HH \rangle + |VV \rangle + |\psi
^+ \rangle \right )\;,  \label{v3}\\
& & |v_{II} \rangle = \frac {1}{\sqrt 3} \left (|HH \rangle + e^{2\pi i/3}|VV
\rangle + e^{-2\pi i /3}|\psi ^+ \rangle \right )\;,  \nonumber \\
& & |v_{III} \rangle = \frac {1}{\sqrt 3} \left (|HH \rangle + e^{-2\pi i/3}|VV
\rangle + e^{2\pi i /3}|\psi ^+ \rangle \right )\;.\nonumber 
\end{eqnarray}
It is quite easy to generate the states of the first basis. On the
other hand, the states of the second basis can be generated according
to the above derivation.

Explicitly one has 
\begin{eqnarray}
|v_{i}\rangle =(U_{i}\otimes W_{i})|\chi \rangle \;,
\label{uvchi}
\end{eqnarray}
where the seed state $|\chi \rangle $---the same for all
i=I,II,III---is written
\begin{eqnarray}
|\chi \rangle  &=&\frac{\sqrt{2}+1}{\sqrt{6}}|HH\rangle +\frac{\sqrt{2}-1}{
\sqrt{6}}|VV\rangle   \nonumber \\
&\simeq &0.986|HH\rangle +0.169|VV\rangle \;,
\end{eqnarray}
and the set of unitaries is given by 
\begin{eqnarray}
&&
U_{I}=W_{I}=\frac{1}{\sqrt{2}}\left( 
\begin{array}{cc}
1 & 1 \\ 
1 & -1
\end{array}
\right) \;;
\nonumber \\& & 
U_{II}=W_{II}=\frac{1}{\sqrt{2}}\left( 
\begin{array}{cc}
1 & 1 \\ 
e^{-2i\pi /3} & e^{i\pi /3}
\end{array}
\right) \;;
\nonumber \\& & 
U_{III}=W_{III}=\frac{1}{\sqrt{2}}\left( 
\begin{array}{cc}
1 & 1 \\ 
e^{2i\pi /3} & e^{-i\pi /3}
\end{array}
\right) \;. 
\end{eqnarray}

Notice that for the particular chosen basis, the unitaries $U_{i}$ and
$W_{i} $ are identical. Moreover, using the factorization formula
(\ref{fact}), the phase-shift on the right reduces to the identity
matrix, and hence the $U_{i}$'s can be implemented by a $\lambda /2$
wp rotated by $\theta =\frac{\pi }{8}$, followed by a further phase
delay between $H$ and $V$, corresponding to $ \varphi _{I}=0$,
$\varphi _{II}=\frac{2}{3}\pi $ and $\varphi _{III}=-\frac{2 }{3}\pi
$, respectively.

\par Other qutrits of the form 
\begin{eqnarray}
|\xi \rangle = \frac{1}{\sqrt 3} (|HH \rangle + e^{i\psi }|VV \rangle
 + e^{i\phi } |\psi ^+ \rangle )
\; 
\end{eqnarray}
can be generated by using the general formula $UDW^\tau =\xi$, where  
\begin{eqnarray}
&&\!\!\!\!\!\!\!\!
U=\frac{1}{\sqrt 2}\left( 
\begin{array}{ll}
e^{i \arg [\sqrt 2 +e^{i(\phi -\frac \psi 2)}]} 
& 
e^{i \arg [\sqrt 2 - e^{i(\phi -\frac \psi 2)}]} 
\\ 
e^{i \arg [e^{i \phi } + \sqrt{2}e^{i \frac \psi 2}]}  
& 
e^{i \arg [e^{i \phi } - \sqrt{2}e^{i\frac \psi 2}]}  
\end{array}
\right) 
\;;
\nonumber \\& & 
\!\!\!\!\!\!\!\!
W=\frac{1}{\sqrt 2}\left( 
\begin{array}{cc}
1
& 
1
\\ 
e^{i \frac \psi 2} 
& 
- e^{i \frac \psi 2} 
\end{array}
\right) 
\;;\nonumber \\
&&\!\!\!\!\!\!\!\!
D=
\left( 
\begin{array}{cc}
\sqrt{\frac 12 +\frac{\sqrt 2}{3}\cos (\frac \psi 2 -\phi)}
& 
0
\\ 
0
& 
\sqrt{\frac 12 - \frac{\sqrt 2}{3}\cos (\frac \psi 2 -\phi)}
\end{array}
\right) 
\;;
\nonumber \\& & 
\!\!\!\!\!\!\!\!
\xi=\frac{1}{\sqrt 3}
\left( 
\begin{array}{cc}
1
& 
\frac{e^{i\phi}}{\sqrt 2}
\\ 
\frac{e^{i\phi}}{\sqrt 2}
& 
e^{i\psi }
\end{array}
\right) 
\;.\label{uv}
\end{eqnarray}
\par Hence, one can write a relation as in Eq. (\ref{uvchi}) for two
other bases which are mutually unbiased with each other and with those
of Eqs. (\ref{u3}) and (\ref{v3}), with the seed state
\begin{eqnarray}
|\chi \rangle &=&\sqrt{\frac{3+\sqrt{2}}{6}}|HH\rangle
+\sqrt{\frac{3-\sqrt{ 2}}{6}}|VV\rangle \nonumber \\ &\simeq
      &0.858|HH\rangle +0.514|VV\rangle \;,
\end{eqnarray}
and with unitaries that can be evaluated by Eq. (\ref{uv}).

The above described procedure could produce highly pure qudits. In
fact, both the technique used to generate non-maximally entangled
states and the waveplates and phase-shifters to realize the unitary
operations can be very accurate, and in principle do not introduce any
amount of mixedness in the state.  Indeed, non-maximally entangled
states generated by this source were recently used to prove the
Hardy's ladder theorem on nonlocality up to the $20th$ step of the
ladder \cite{roma3}.

Once qudits are available, one can characterize these states by
quantum tomography, or use them for more advanced tests of nonlocality
\cite{collins:02,zuk}. As far as more specific quantum information
applications are concerned, e.g. quantum key distribution, a major
difficulty is the need to perform quantum measurements on mutually
unbiased bases.  The use of qutrits requires highly nontrivial setups
at the measurement stage. However, the use of ququads is easier.  In
this case one should use five mutually unbiased bases, hence
generating $20$ different states. For a system of two qubits
\cite{zeil,arav}, one can consider three product bases and two Bell
bases. We write explicitly the bases from Ref. \cite{arav} (in our
scheme, we have $|0\rangle \equiv |H\rangle $ and $|1\rangle \equiv
|V\rangle $), namely
\begin{eqnarray}
I)\ \  &&|0\rangle |0\rangle ,|0\rangle |1\rangle ,|1\rangle |0\rangle
,|1\rangle |1\rangle   \;;
\nonumber \\
II)\ \  &&(|0\rangle +|1\rangle )(|0\rangle \pm |1\rangle ),  \nonumber \\
&&(|0\rangle -|1\rangle )(|0\rangle \pm |1\rangle )  \;;\nonumber \\
III)\ \  &&(|0\rangle +i|1\rangle )(|0\rangle \pm i|1\rangle ),  \nonumber \\
&&(|0\rangle -i|1\rangle )(|0\rangle \pm i|1\rangle )  \;;\nonumber \\
IV)\ \  &&(|0\rangle +i|1\rangle )|0\rangle \pm (|0\rangle -i|1\rangle
)|1\rangle ,  \nonumber \\
&&(|0\rangle -i|1\rangle )|0\rangle \pm (|0\rangle +i|1\rangle )|1\rangle \;; 
\nonumber \\
V)\ \  &&|0\rangle (|0\rangle +i|1\rangle )\pm |1\rangle (|0\rangle
-i|1\rangle ),  \nonumber \\
&&|0\rangle (|0\rangle -i|1\rangle )\pm |1\rangle (|0\rangle +i|1\rangle )\;.
\end{eqnarray}
Clearly, the bases $I$,$II$, and $III$ correspond to the measurement
of $ \sigma _{z}\otimes \sigma _{z}\;,\sigma _{x}\otimes \sigma _{x}$,
and $ \sigma _{y}\otimes \sigma _{y}$, respectively. The bases $IV$
and $V$ are made of Bell projectors. The generation of the $12$
product states is trivial. On the other hand, the above source of
entangled photon states very efficiently generates the other eight
maximally entangled states.

\par The problem of realizing Bell measurements can be solved by
hyper-entangled states \cite{padua}, which have been realized in the
two degrees of polarization and momentum by the same source
\cite{roma2}. Besides polarization entanglement, momentum entanglement
is realized by a four hole screen which allows one to select the
correlated pairs of modes $a_{1},b_{2}$ and $a_{2},b_{1}$ (Fig. 1)
occurring with relative phase $\phi =0$. In this way, in either one of
the cones the momentum entangled Bell state $|\psi ^{+}\rangle
=\frac{1}{\sqrt{2 }}\left( |a_{1},b_{2}\rangle +|b_{1},a_{2}\rangle
\right) $ can be generated. Note that the four modes $ a_{1}$,
$b_{2}$, $a_{2}$, $b_{1}$ can be easily separated by mirrors and
coupled to single mode optical fibers, allowing in this way fiber
based cryptographic schemes. In a complete Bell state analysis the
polarization state acts as the control qubit and the momentum state
$|\psi ^{+}\rangle $ as the target qubit \cite{padua}.

\par We notice also that a cryptographic protocol with ququads that
uses just two instead of all five mutually unbiased bases is
characterized by a maximum acceptable error rate that is only slightly
lower, while the corresponding key rate is much larger \cite{eav}. The
nontrivial encoding here is represented by the Bell states of the
bases $IV$ and $V$, and a cryptographic scheme based just on such two
bases can be implemented by our source.

\par In conclusion, we have shown how to obtain an arbitrary qudit up
to $d=4$, from local unitary transformations on a bipartite pure state
of two qubits by SVD encoding.  The theoretical scheme generally holds
in arbitrary Hilbert space, encoding the qudit on a bipartite quantum
system of ${\cal H}\otimes {\cal H}$ by means of local unitaries, with
$d= (\hbox{dim}({\cal H}))^2$.  Upon representing qubits by the
polarization state of photons, the method allows one to generate
experimentally qudits with a single nonlinear crystal and linear
optics, using the source of Ref. \cite{roma1}.  This allows one to create
tunable nonmaximally entangled states that play the role of seed
states, from which arbitrary qudit states are generated via SVD using
simple linear optics. The hyper-entanglement of the generated photons
allows one to perform nontrivial measurements---such as Bell
measurements---that are crucial for quantum cryptographic
applications.

\emph{Acknowledgments.} This work has been sponsored by INFM through
the project PRA-2002-CLON, and by EC and MIUR through the cosponsored
ATESIT project IST-2000-29681 and Cofinanziamento 2003.

\end{document}